\documentclass[twocolumn,superscriptaddress,showpacs,preprintnumbers,prx,amsmath,amssymb]{revtex4}

\usepackage{graphicx}
\usepackage{bm}
\usepackage{epsfig}
\usepackage{graphicx}
\usepackage{bm}
\usepackage{epsfig}
\usepackage{comment} 

\newcommand{\Sr}{SrFe$_2$As$_2$}

\newcommand{\Ca}{CaFe$_2$As$_2$}
\newcommand{\CaR}{Ca$_{1-x}$R$_x$Fe$_2$As$_2$}

\newcommand{\ie}{{\it i.e.}}
\newcommand{\eg}{{\it e.g.}}
\newcommand{\etal}{{\it et al.}}

\begin{document}

\title{Separation of antiferromagnetism and high-temperature \\
superconductivity in Ca$_{1-x}$La$_x$Fe$_2$As$_2$ under pressure}


\author{Shanta~R.~Saha}
\author{T.~Drye}
 \affiliation{Center for Nanophysics and Advanced Materials, Department of Physics, University of Maryland, College Park, MD 20742}
\author{S.K.~Goh}
\author{L.E.~Klintberg}
\author{J.M.~Silver}
\author{F.M.~Grosche}
\author{M.~Sutherland}
\affiliation{Cavendish Laboratory, University of Cambridge, J.J. Thomson Avenue, Cambridge CB3 0HE, U.K.}
\author{T.J.S. Munsie}
\author{G.M. Luke}
\affiliation{1Department of Physics and Astronomy, McMaster University, 1280 Main Street West, Hamilton, Ontario, Canada L8S 4M1}
\author{D.~K.~Pratt}
\author{J.~W.~Lynn}
\affiliation{NIST Center for Neutron Research, National Institute of Standards and Technology, Gaithersburg, MD 20899}
\author{J.~Paglione}
\affiliation{Center for Nanophysics and Advanced Materials, Department of Physics, University of Maryland, College Park, MD 20742}
\email{paglione@umd.edu}

\date{\today }


\begin{abstract}

We report the effect of applied pressures on magnetic and superconducting order in single
crystals of the aliovalent La-doped iron pnictide material Ca$_{1-x}$La$_{x}$Fe$_{2}$As$_{2}$. Using electrical transport, elastic neutron scattering and resonant tunnel diode oscillator measurements on samples under both quasi-hydrostatic and hydrostatic pressure conditions, we report a series of phase diagrams spanning the range of substitution concentrations for both antiferromagnetic and superconducting ground states that include pressure-tuning through the antiferromagnetic (AFM) quantum critical point.
Our results indicate that the observed superconducting phase with maximum transition temperature of $T_{c}$=47~K is intrinsic to these materials, appearing only upon suppression of magnetic order by pressure tuning through the AFM critical point. In contrast to all other intermetallic iron-pnictide superconductors with the ThCr$_2$Si$_2$ structure, this superconducting phase appears to exist only exclusively from the antiferromagnetic phase in a manner similar to the oxygen- and fluorine-based iron-pnictide superconductors with the highest transition temperatures reported to date. The unusual dichotomy between lower-$T_{c}$ systems with coexistent superconductivity and magnetism and the tendency for the highest-$T_{c}$ systems to show non-coexistence provides an important insight into the distinct transition temperature limits in different members of the iron-based superconductor family.

\end{abstract}

\maketitle

\section{Introduction}

The interplay between structural, magnetic and superconducting properties in the newly discovered iron-based superconducting compounds has been a central theme in attempts to elucidate the nature of Cooper pairing in this new family of high-temperature superconductors~\cite{Kamihara,Paglione,Johnston,Stewart}. In particular, manipulation of the electronic structure via chemical substitution or applied pressure is thought to play a key role in the disruption of antiferromagnetic order and the stabilization of superconductivity~\cite{Paglione}. 
These similar but unique tuning parameters generically produce the well-known phase diagram of the iron-based superconductors, with an antiferromagnetic (AFM) ordering temperature $T_N$ continuously suppressed toward zero as a function of doping, substitution or pressure, before impinging on a superconducting (SC) state with transition temperature $T_c$ that also varies as a function of tuning parameter. The interplay of these two phases has been the focus of much efforts in understanding the nature and origin of both ground states. 

Experiments have shown a variety of behavior including both exclusive separation of the AFM and SC phases as well as microscopic coexistence of the two. As first observed in the RFeAsO ``1111'' oxypnictide compounds~\cite{Ren1}, which continue to possess the highest $T_c$ values in the entire family, the AFM and SC phases appear to show little to no coexistence as observed in LaFeAsO$_{1-x}$F$_x$~\cite{Luetkens}, CeFeAsO$_{1-x}$F$_x$~\cite{Zhao}, and SmFeAsO$_{1-x}$F$_x$~\cite{Maeter} systems, even exhibiting a first-order-like transition between the two phases. In contrast, the intermetallic ``122'' compounds with the ThCr$_2$Si$_2$ crystal structure - such as BaFe$_{2-x}$Co$_x$As$_2$ \cite{Ni,Chu,Klintberg,Pratt-PRL} and Ba$_{0.77}$K$_{0.23}$Fe$_2$As$_2$~\cite{Li-PRB} - exhibit a wide range of microscopic coexistence between AFM and SC phases which has even been claimed at the atomic level~\cite{Laplace-PRB80}. While mean-field theories have proven useful in explaining the cooperative competition between AFM and SC transitions and the manner by which the two phases shape the phase diagram of the 122 materials~\cite{Fernandes-PRB81, Fernandes-PRB82, Fernandes-PRL}, the exclusive repulsion of the two phases in the 1111 materials and its relevance to achieving the highest $T_c$ values remains to be understood \cite{Goremychkin,Fujiwara}. 

In the smallest unit cell member of the 122 family, CaFe$_{2}$As$_{2}$, AFM order can be suppressed by both chemical substitution (\eg, Co and Ni for Fe~\cite{Kumar-Co,Kumar-Ni}, Na for Ca~\cite{Zhao-Na}) and applied pressure \cite{Alireza,Torikachvili,Park,Yu} to reveal superconductivity in a manner much like the other 122 systems. Recently, we have reported a new approach to tuning the CaFe$_{2}$As$_{2}$ phase diagram using rare earth substitution, utilizing the doping effect of replacing divalent Ca$^{2+}$ with trivalent R$^{3+}$ rare earth elements (La, Ce, Pr, and Nd) in Ca$_{1-x}$R$_x$Fe$_{2}$As$_{2}$~\cite{Saha_CaR}, as well as a chemical pressure effect due to the ionic size mismatch between Ca and the light rare earths. This approach induces superconductivity with the highest transition temperatures yet reported for the intermetallic iron-based superconductors~\cite{Rotter,Akimitsu}, with $T_c$ values reaching as high as 49~K \cite{Saha_CaR,Gao,Lv,Drye}.
However, the lack of a full-volume-fraction SC phase in all R-substituted series has raised questions about the intrinsic nature of superconductivity in this system, which appears in both uncollapsed (for La and Ce) and collapsed (for Pr and Nd) tetragonal states~\cite{Saha_CaR}. While the former may be aleviated by recent reports of stronger Meissner screening in phosphorus-substituted samples~\cite{Kudo}, the latter case is particularly intriguing since the widely believed spin fluctuation pairing mechanism is likely inactive in the collapsed state due to the quenched iron moment \cite{Yildirim,Goldman,Gretarsson,Ma}.

Here we report the intrinsic separation of AFM and SC phases in the La-doping series, Ca$_{1-x}$La$_x$Fe$_{2}$As$_{2}$, found by fine-tuning samples with a range of La concentrations through the AFM-SC critical point with applied pressure. We performed electrical resistivity and neutron scattering experiments to investigate this evolution as a function of pressure, utilizing both quasi-hydrostatic as well as hydrostatic pressure experiments to rule out extrinsic strain effects. With clear evidence of an onset of the superconducting phase upon continuous pressure tuning beyond the AFM ordered phase, we conclude that the superconductivity observed in rare earth-doped \Ca\ is intrinsic to the material and does not arise due to an impurity phase. With a resultant phase diagram strikingly similar to that of the 1111 pnictide superconductors, we draw parallels between the two systems and speculate on a common origin of superconductivity with the highest transition temperatures in the iron-based family.
Experimental details and organization of the pressure experiments are outlined in Section II, followed by discussion of our main results for each pressure experiment in Section III and general conclusions in Section IV.

\section{Experimental Details}

Single-crystal samples of Ca$_{1-x}$La$_{x}$Fe$_{2}$As$_{2}$~ were grown using the FeAs self-flux method,~\cite{Saha_Sr} yielding crystals as large as $\sim 10\times 10\times $ 0.1 mm$^{3}$. Chemical analysis was obtained via both energy-dispersive (EDS) and wavelength-dispersive (WDS) x-ray spectroscopy, showing 1:2:2 stoichiometry between (Ca,La), Fe, and As concentrations. The actual La concentration was determined using WDS and single-crystal x-ray diffraction measurements~\cite{Saha_CaR}.

\begin{figure}[t]
\centering
\resizebox{8cm}{!}{
  \includegraphics[width=8cm]{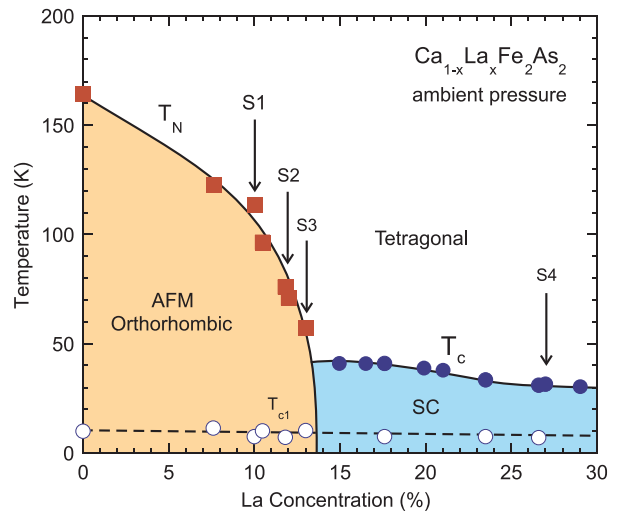}}
\caption{\label{fig1} Lanthanum substitution phase diagram of Ca$_{1-x}$La$_{x}$Fe$_{2}$As$_{2}$, showing the antiferromagnetic (AFM) transitions at $T_N$
~(squares) and superconducting transitions (circles). Open circles indicate the ``10~K'' superconducting transition $T_{c1}$ (see text). Arrows indicate the samples S1, S2, S3, and S4 (with La concentrations $x$=0.10, 0.12, 0.13 and 0.27, respectively) investigated with applied pressure experiments in this study.}
\end{figure}

Fig.~\ref{fig1} presents the La substitution phase diagram for Ca$_{1-x}$La$_{x}$Fe$_{2}$As$_{2}$, showing the evolution of AFM and SC transition temperatures as determined by magnetic susceptibility and electrical resistivity measurements, respectively. For details we refer to our previous study~\cite{Saha_CaR}. As shown, La substitution suppresses AFM order and induces superconductivity with $T_c$ values reaching 40~K in concentrations above 14\%. Throughout the range of La substitutions a second superconducting transition with $T_c \simeq 10$~K is always present in trace forms. This phase is believed to be related to the filamentary superconducting phase~\cite{Xiao-PRB85} induced in \Ca\ under quasi-hydrostatic pressure conditions (\ie, a typical liquid-medium clamp cell pressure experiment) ~\cite{Torikachvili,Alireza} that is absent under true hydrostatic pressure conditions (\ie, in a helium gas cell) \cite{Yu}. In the present study we focus on the high-$T_c$ phase that occurs with $T_c$ values exceeding 30~K, which does not show any traces in samples with $x<0.14$ under ambient conditions and is believed to be of different origin. We have chosen three ``underdoped'' (\ie, $x<0.14$) samples and one ``overdoped'' (\ie, $x>0.14$) sample to investigate under pressure, indicated by the positions of the arrows in Fig ~\ref{fig1}. 

The experiments are organized as follows.
First, electrical resistivity measurements under quasi-hydrostatic pressure were carried out at the University of Maryland using the four-wire contact method on sample S1 ($x$=0.10, $T_{N}(0)$=114~K), employing a miniature piston-cylinder cell with maximum pressure range of 15~kbar. The sample was loaded in a Fluorinert liquid presure medium along with a tin manometer. Increasing pressure was applied systematically after each thermal cycle, with resistance measurements taken on cooling. Pressures were calibrated at low temperature by measuring the resistance of the tin manometer and using the change of its superconducting transition temperature with pressure.

Second, hydrostatic pressure experiments were carried out using helium gas pressure systems. Compared with the clamp cell experiment, which employs a liquid pressure medium that solidifies at low temperatures, the use of helium in a gas cell provides the best possible hydrostatic pressure sample enviromnent over a much wider range of temperatures and pressures. In particular, the He freezing point only increases to about 50 K at $P$=0.7 GPa (7 kbar), allowing the best possible hydrostatic sample environment in our experimental setup. 

Neutron diffraction studies under He gas pressure were carried out on sample S2 ($x$=0.12, $T_{N}(0)$=70~K) using the BT-4 Filter Analyzer Triple-Axis Spectrometer at the NIST Center for Neutron Research using an Al-alloy He-gas pressure cell to ensure hydrostatic pressure conditions, as reported previously~\cite{Goldman}. The cell was connected to a pressurizing intensifier through a high-pressure capillary that allowed continuous monitoring and adjusting of the pressure. Using this system, the pressure could be varied at fixed temperatures above the He solidification line or temperature could be scanned at nearly constant pressures. A helium reservoir allowed the pressure to remain relatively constant as the temperature was changed.

Electrical resistivity measurements under He gas pressure were carried out at McMaster University on sample S3 ($x$=0.13, $T_{N}(0)$=58~K), measured using a helium gas pressure system as reported previously \cite{Yu}. The sample was loaded in the pressure cell with a standard four-probe configuration, and cooled in a helium storage dewar, with pressure applied in-situ by an external helium compressor. The maximum pressure attained was 4~kbar for the helium pressure cell, and we employed a cooling rate of approximately 1 K/min through the structural transitions. To ensure that the applied pressure in the compressor was transmitted to the pressure cells, we always set the pressure while the system temperature was maintained above 70 K, thus avoiding the possibility of helium freezing in the feed capillary. 

Finally, higher pressures were achieved using Moissanite anvil cell experiments to study superconductivity in sample S4 ($x$=0.27, $T_{c}(0)$=31~K) at Cambridge University, both by conventional four-wire resistivity measurements and in a similar sample by tracking the resonant frequency of an oscillator formed by a tunnel diode and a micro-coil in the gasket hole of a Moissanite anvil cell \cite{Goh}. Ruby fluorescence spectroscopy was used for pressure determination, and glycerin was used as the pressure transmitting fluid.


\section{Experimental Results}

\subsection{Quasi-Hydrostatic Clamp Cell}

\begin{figure}[!t]
\centering
\resizebox{8cm}{!}{
  \includegraphics[width=8cm]{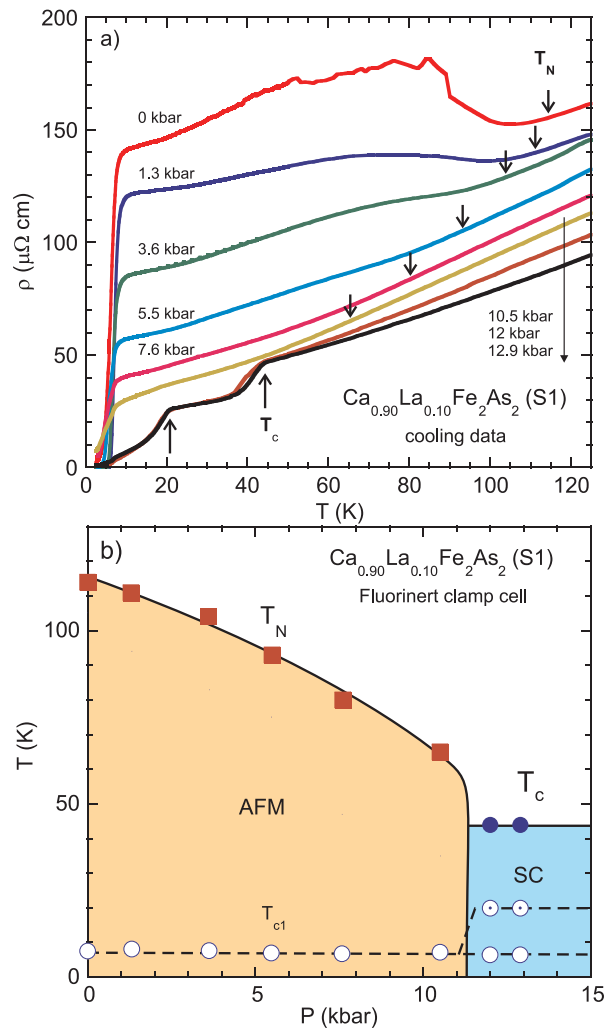}}
\caption{\label{fig2} Quasi-hydrostatic pressure experiment on Ca$_{0.9}$La$_{0.1}$Fe$_{2}$As$_{2}$ (underdoped sample `S1' in Fig.~\ref{fig1}) using a piston-cylinder clamp cell with sample loaded in a Fluorinert liquid presure medium. 
(a) Resistivity measured upon cooling temperature as a function of applied pressures. Down arrows indicate position of the antiferromagnetic ordering temperatures $T_N$ as defined by the kink in $d\rho/dT$ (see text and Supplementary Material \cite{SUPP}). Up arrows indicate the onset of superconducting transitions, including the high-$T_{c}$ phase  at 44~K, that appear abruptly upon increase of pressure from 10.5~kbar to 12~kbar.
b) Phase diagram for sample S1, showing the evolution of $T_N$ (squares) and superconducting transition temperatures $T_{c}$ (closed circles) and $T_{c1}$ (open circles), which is related to the ``10~K'' phase observed in undoped CaFe$_{2}$As$_{2}$ at ambient pressure. An intermediate superconducting phase appears above 10.5~kbar near $\sim $20~K (dotted circles).}
\end{figure}

Resistivity measurements of sample S1 ($x$=0.10, $T_{N}(0)$=114~K) were carried out under quasi-hydrostatic pressures with a piston-cylinder clamp cell using liquid (Fluorinert) pressure medium, and are presented in Fig.~\ref{fig2}(a). At ambient pressure, the AFM transition is indicated by the upward rise of resistivity $\rho$ with decreasing temperature that results in a kink in the temperature derivative $d\rho/dT$ that coincides with the onset of a step in the magnetic susceptibility at $T_{N}(0)$=114~K (see Supplementary Material \cite{SUPP}).
This value of $T_{N}(0)$=114~K, suppressed from 165~K in the parent compound, allows for a more accessible range of behavior to be reached in a conventional piston-cylinder clamp cell. Upon application of pressure, the overall resistivity values are suppressed, along with the AFM transition temperatures. Because the distinct onset of AFM order is smeared out as pressure is increased, we define $T_{N}(P)$ (\ie, the pressure-dependent $T_N$) to be the temperature below which $d\rho/dT$ deviates from a constant value. This translates to the onset of curvature in the resistivity curves (see Supplementary Material \cite{SUPP}), as indicated by the position of the arrows in Fig.~\ref{fig2}a). 
The lowest detectable trace of $T_N$ is at $\sim$65~K at 10.5~kbar, yielding a pressure coefficient of $dT_N$/dP$\simeq -4.7$~K/kbar for the range $T_N(P)$ between 114~K and 65~K. This evolution is shown in the pressure-temperature phase diagram presented in Fig.~\ref{fig2}(b).

At low temperatures, a superconducting transition at $T_{c1}\simeq 8$~K, believed to be related to the ``10~K'' phase noted above, is observable at ambient pressure and does not appear to evolve with pressure, as shown in the phase diagram in Fig.~\ref{fig2}b). Increasing pressure to a higher value reveals two new features that suddenly emerge as abrupt drops in resistivity at $T_{c}=$44~K and $T_{c2}=$20~K. The dramatic appearance of these features appear to be associated with the suppression of AFM order, as shown in Fig.~\ref{fig2}b): once AFM order is suppressed below the value of the high transition temperature of 44~K, superconductivity emerges. While the transitions are not complete, the trace of superconductivity is believed to be associated with the same phase that is induced with La substitution as shown in Fig.~\ref{fig1} and with other rare earth substitutions \cite{Saha_CaR}, suggesting pressure and doping are acting to produce very similar phase diagrams in this range of parameters.
The reason for the appearance of two partial transitions (\ie, at 20~K and 44~K) is not clear.  Upon close inspection of the resistivity data, the signature of $T_{c1}$ near 8~K is still present above the critical pressure where $T_N$ is suppressed, indicating its presence is impervious to any change in the overall ground state of the material. In contrast, the signatures of superconductivity at 20~K and 44~K abruptly appear beyond the quantum phase transition. While we believe the 44~K transition to be unique to the \CaR\ series, the 20~K phase may be related to another strain-induced phase that is well documented in the case of \Sr\ \cite{Saha_Sr} to be driven by lattice distortions induced by strained sample conditions.

\begin{figure}[!t]
\centering
\resizebox{8cm}{!}{
  \includegraphics[width=8cm]{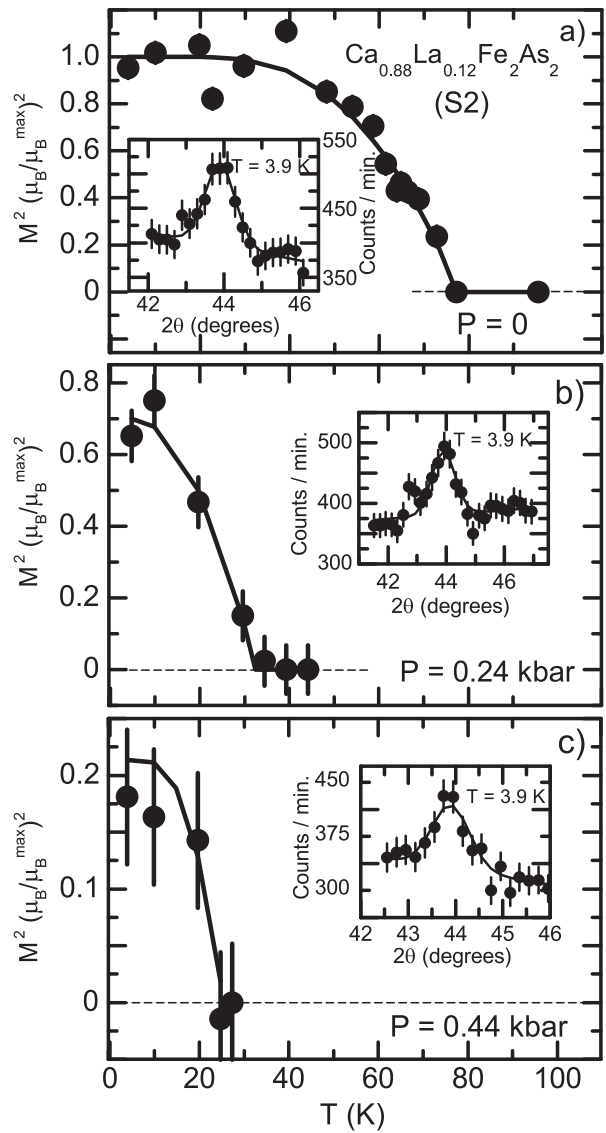}}
\caption{ \label{fig3} Hydrostatic pressure dependence of ordered AFM moments in Ca$_{0.88}$La$_{0.12}$Fe$_{2}$As$_{2}$~ (underdoped sample `S2' in Fig. 1) determined by neutron scattering experiments on the BT-4 beam line at the NCNR. Integrated intensities measured at the $Q_{AFM}$=(1 0 3) magnetic Bragg peak are shown for selected applied pressures of a) $P$=0, b) $P$=0.24~kbar and c) $P$=0.44~kbar. Intensities are proportional to the square of the magnetic order parameter $M^2$, which is the ordered moment at $(P,T)$.  Values are normalized to the maximum ambient pressure value observed at 3.9~K.  Solid lines are mean field order fits to $M^2$ to obtain the N\'eel temperatures for each pressure.  Insets show data from $\theta$-2$\theta$ scans at 3.9 K, fit to a Guassian line shape with a sloping background.}
\end{figure}

Because relatively high pressures are needed to achieve the range where these transitions appear, it is possible that a transition to the collapsed tetragonal (cT) phase occurs in this region. 
Nevertheless, the sudden appearance of superconductivity with the loss of AFM order in a pressure window of less than 2~kbar indicates the mutually exclusive nature of the two phases. This is strikingly similar to that observed in 1111 iron-based superconductors, where the two phases compete for the ground state~\cite{Luetkens,Zhao}. Furthermore, it also confirms that the trace high-$T_c$ phase is highly unlikely to be extrinsic in origin, as recently suggested to originate from structural clustering \cite{Lv2} that would presumably exhibit a signature in resistivity at all pressures and not onset upon pressure increase, as observed here. Rather, this must be an intrinsic phenomenon stabilized by rare earth substitution and/or the suppression of AFM order. The role of non-hydrostaticity on this phenomenon is considered next.

\subsection{Hydrostatic Helium Cell}

\subsubsection{Neutron Experiment}

\begin{figure}[!t]
\centering
\resizebox{8cm}{!}{
  \includegraphics[width=8cm]{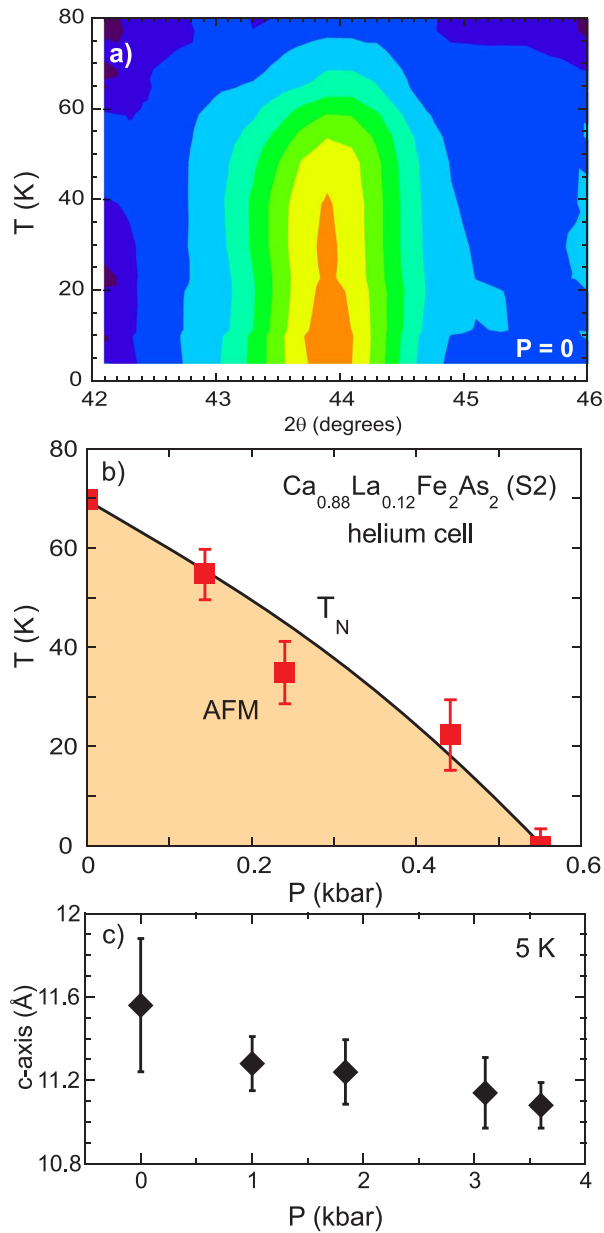}}
\caption{ \label{fig32} 
(a) Magnetic intensity of the (1~0~3) Bragg peak at ambient pressure
in Ca$_{0.88}$La$_{0.12}$Fe$_{2}$As$_{2}$ (underdoped sample `S2' in Fig. 1)
determined from neutron diffraction on the BT-4 beam line at the NCNR.
(b) Hydrostatic pressure dependence of $T_N$~ in Ca$_{0.88}$La$_{0.12}$Fe$_{2}$As$_{2}$~  determined by magnetic order parameter fits (see Fig.~\ref{fig3}). 
(c) Variation of the $c$-axis lattice parameter vs. pressure determined by measuring the (0~0~4) structural Bragg reflection at 5~K. Uncertainties are statistical in origin and represent one standard deviation.}
\end{figure}

In order to further clarify the interplay of AFM and high-$T_{c}$ superconductivity, we investigated the magnetic properties of another underdoped sample with a lower $T_N(0)$ (\ie, closer to the border of the SC dome) under hydrostatic pressure in a helium pressure cell. 
Neutron diffraction experiments were carried out on sample S2 ($x$=0.12, $T_{N}(0)$=70~K) under applied He gas pressure, allowing determination of both the magnetic and structural phases directly.
Fig.~\ref{fig3} presents the normalized data and fits to the integrated intensities for a few selected pressures.  An example of the scattering intensity as a function of temperature and diffraction angle is presented as a contour plot in Fig.~\ref{fig32}a). This intensity is proportional to the square of the staggered magnetization, or AFM order parameter, as plotted in Fig.~\ref{fig3} along with mean field fits. The resultant fit estimates of $T_N(P)$ are summarized in Fig.~\ref{fig32}a) as a function of pressure.  The value of $T_{N}(0)$=70~K at ambient pressure (Fig.~\ref{fig3}a) is consistent with that determined by the drop in the magnetic susceptibility (not shown; see Supplementary Material \cite{SUPP}).

As shown in Fig.~\ref{fig3}, the value of $T_{N}(P)$ is dramatically suppressed with increasing hydrostatic pressure, with no magnetic peak detectable at 0.55~kbar pressure down to 4~K (not shown). This rate of suppression, $dT_N/dP \simeq$-125~K/kbar, is profoundly different than that found in the quasi-hydrostatic clamp cell experiment presented in Fig.~\ref{fig2}, where $dT_N/dP \simeq$-4.7~K/kbar is approximately 25 times smaller. It is in fact comparable to the largest reported pressure coefficient $dT_N/dP \simeq$-110~K/kbar reported in Ca(Fe$_{1-x}$Co$_{x}$)$_{2}$As$_{2}$ under hydrostatic pressure~\cite{Gati}, and is  

The reason for such dramatic differences in pressure dependence is debatable. It is certainly likely that $dT_N/dP$ is a nonlinear parameter that is strongly dependent on the rare earth concentration (and thus depends very sensitively on small changes in lattice parameters, since La substitution does not introduce significant changes to the unit cell \cite{Saha_CaR}). The degree of pressure homogeneity is certainly a factor, and the Poisson effect due to anisotropic change of lattice parameters under inhomogenous pressure combined with an anisotropic pressure transmission from a frozen pressure medium could possibly lead to such differences. In any case, it is well established that the \Ca\ system is extremely sensitive to the nature of structural perturbation caused by chemical or applied pressure application due its proximity to a structural instability that drives a change in the bonding structure in the ThCr$_2$Si$_2$-type unit cell. The proximity of interlayer As-As dimer formation to the range of unit cell parameters encountered upon moderate tuning of CaFe$_{2}$As$_{2}$ changes the bonding structure of the crystal in a very abrupt structural collapse of the unit cell \cite{Kreyssig,Goldman,Saha_CaR} that occurs midway through the AFM-SC phase diagram of the undoped system, amplifying the effects of strain induced by non-hydrostatic pressure conditions. 


Neutron and x-ray scattering and transport studies in CaFe$_{2}$As$_{2}$ under hydrostatic pressure have shown that $T_N$ is rapidly suppressed and at pressure of 3.5~kbar the system undergoes another dramatic structural phase transition into the cT phase \cite{Kreyssig,Goldman}. 
In order to determine the possible proximity of the pressure-induced collapse transition in our pressure experiments, measurements of the $c$-axis lattice parameter were obtained up to 3.6~kbar at 5~K. As shown in Fig.~\ref{fig3}(c), the variation of the $c$-axis lattice parameter with pressure exhibits no significant change with pressure, demonstrating the absence of any structural collapse up to at least $\sim $4~kbar. However, the obvious sensitivity of both the magnetic and structural stability of the \CaR\ system to the nature of applied strain suggests that all physical properties, including the appearance of superconducting phases, are strongly coupled to the delicate structural instabilities in this system.

\subsubsection{Resistivity Experiment}

\begin{figure}[!t]
\centering
\resizebox{8cm}{!}{
  \includegraphics[width=8cm]{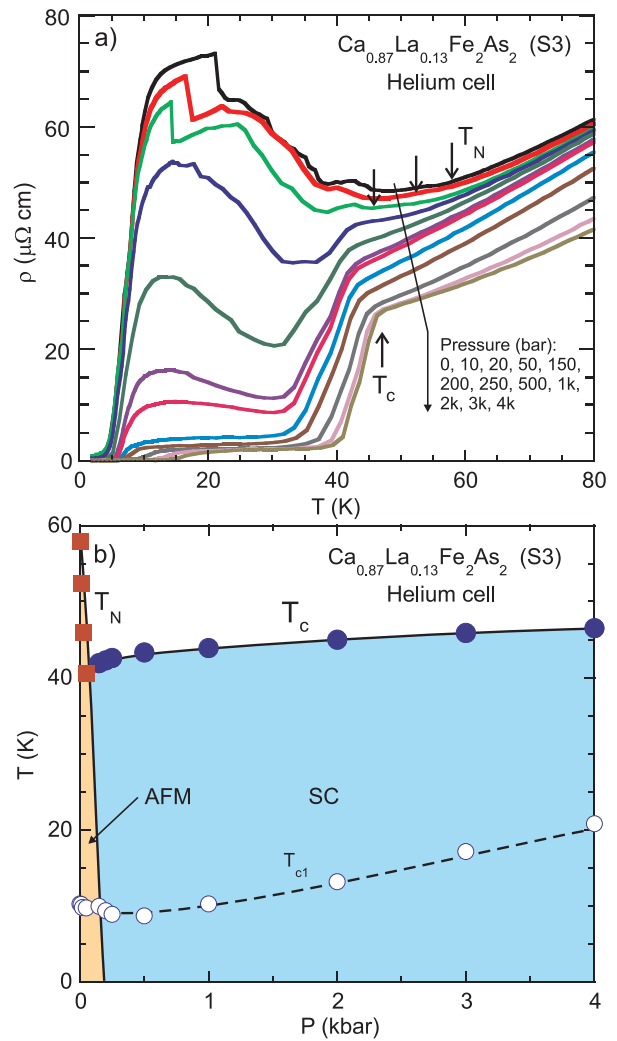}}
\caption{\label{fig4} Hydrostatic pressure dependence of electrical resistivity of Ca$_{0.87}$La$_{0.13}$Fe$_{2}$As$_{2}$ (underdoped sample `S3' in Fig.~\ref{fig1}) using helium gas as the pressure medium. 
(a) Resistivity temperature dependence as a function of applied pressures. Down arrows mark the antiferromagnetic ordering transition temperature $T_N$ for each pressure, as determined by $d\rho/dT$ analysis (see text), and the up arrow indicates an example of the onset of superconductivity at $T_{c}$ for the sample under 4~kbar pressure.
(b)Temperature vs. hydrostatic pressure phase diagram of antiferromagnetic (squares) and superconducting transition temperatures $T_c$ (solid circles) and $T_{c1}$ (open circles).}
\end{figure}

In a separate He-gas pressure experiment, resistivity of a sample even closer to the SC-AFM boundary was measured in a hydrostatic helium cell at McMaster University. Figure \ref{fig4}a) presents the temperature dependence of the resistance of sample S3 ($x$=0.13, $T_{N}(0)$=58~K) for a series of applied pressures. 
At ambient pressure, the resistivity of this sample exhibits an upturn upon decreasing temperature at 58~K, as determined by a change in $d\rho/dT$ (see Supplementary Material \cite{SUPP}); we use this to define $T_{N}(0)$=58~K as the AFM transition (consistent with that used above for sample S1). At lower temperatures, there is a sharp drop of resistance due to the ``10 K'' superconducting transition at T$_{c1}$=10~K, as also observed in sample S1 (see Fig\ref{fig2}). With increasing pressure $T_N$~ shifts to lower temperatures as indicated by the positions of the down arrows, demonstrating a rapid suppression of magnetic order. At only 50~bar, $T_N$ decreases to $\sim$40~K, yielding a pressure coefficient $dT_N/dP \sim -350$~K/kbar. This is even more dramatic than that found in the neutron experiment above for sample S2, where $dT_N/dP \simeq$-125~K/kbar, suggesting a non-linear dependence of the rate of change of $T_N$ with pressure on rare earth concentration is certainly present.

At $P$=50~bar, an abrupt drop in resistance begins to appear, consistent with the onset of high-$T_{c}$ superconductivity at 40~K.  Upon further pressure increase, the drop develops to an almost complete resistive transition with onset temperature at 46.5~K and no trace of the AFM resistance minimum. 
The results shown in fig.~\ref{fig4} set an upperbound of $\sim$200~bar under hydrostatic pressure for the coexistence between the AFM and high-$T_{c}$~SC phases, if they coexist at all. This high-$T_{c}$~SC phase reaches its maximum $T_{c}$=46.5~K at 4~kbar. The resistivity not reaching zero below the high-$T_{c}$~SC transition indicates partial volume fraction superconductivity as studied previously \cite{Saha_CaR}, which remains unexplained. At the disappearance of $T_N$, the $T_{c1}$ phase is enhanced with pressure, reaching a maximum value of $T_{c1}\simeq $20~K at 4~kbar. 

The incredibly small pressure scales used in this experiment (for instance, beginning at 10~bar or $\sim 10$ atmospheres of pressure) to induce rather large changes in transport properties are strongly suggestive that the application of pressure does not induce any irreversible pressure or strain effect that may lead to an extrinsic phase. Also, note that this transformation cannot originate from a transition to the cT phase, as demonstrated by the neutron experiment lattice parameter study presented in Fig.~\ref{fig3}c). Rather, this abrupt appearance of the high-$T_{c}$ transition coincident with the suppression of AFM order again confirms the intrinsic origin of this superconducting phase and its exclusive existence.

The apparent coexistence of the high-$T_{c}$ superconductivity and the AFM in an extremely narrow pressure window ($\leq $0.2~kbar) is similar to what was also found in the 1111 family of superconductors by $\mu$SR experiments~\cite{Maeter}, which is a very sensitive bulk probe. It should be noted that no apparent coexistence of AFM and superconductivity is discussed in the low-$T_{c}$ $\simeq$ 15~K system Ca(Fe$_{1-x}$Co$_{x}$)$_{2}$As$_{2}$ under hydrostatic pressure, ascribing it to the strong first-order character of the magnetostructural transition \cite{Gati}, but partial volume fraction superconductivity is still seen in the AFM state in that phase diagram.

\subsection{Moissanite Anvil Cell}

\begin{figure}[t]
\centering
\resizebox{8cm}{!}{
  \includegraphics[width=8cm]{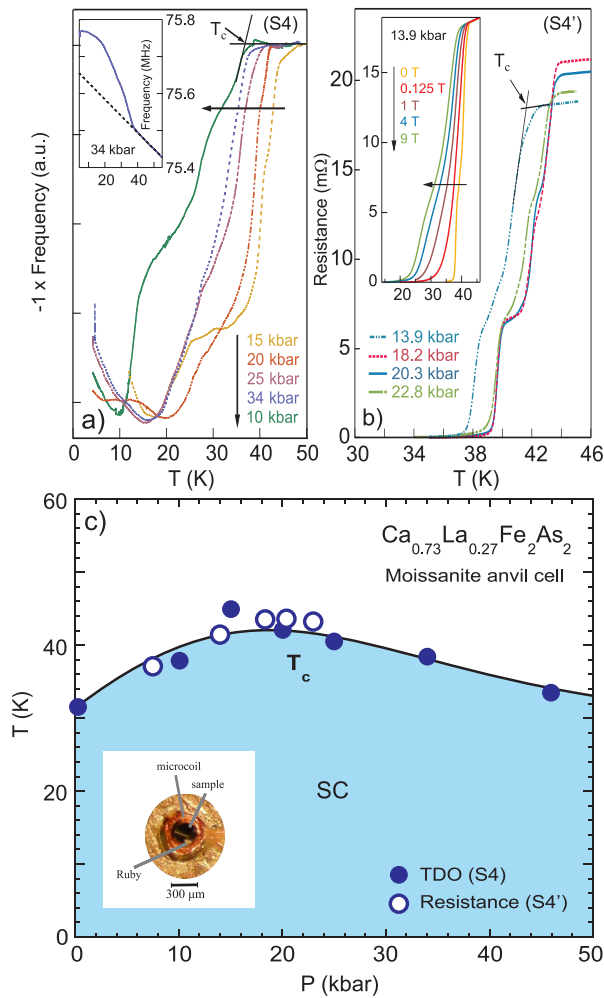}}
\caption{\label{fig5} a) Temperature dependence of the TDO frequency at various pressures
in the Ca$_{0.73}$La$_{0.27}$Fe$_{2}$As$_{2}$ (overdoped sample `S4' in Fig. 1)
single crystal measured in a moissanite anvil cell. Note that the sign of the vertical
axis has been inverted. The upper inset shows the raw data and the
estimation of the background. The lower inset is a photograph of a 10-turn
microcoil with a diameter of 300 $\protect\mu $m enclosing the sample and a
ruby chip. b) Phase diagram showing the pressure dependence of $T_{c}$~
constructed from the resistance (open circles) measured using a
piston-cylinder cell (data shown in supplementary information) and TDO measurements (closed circles).
The solid curve is a guide to the eye.}
\end{figure}

In order to understand the response of superconductivity in ``overdoped'' samples to pressure, we have investigated overdoped sample S4 ($x$=0.27, $T_{c}(0)$=31~K) in the higher quasi-hydrostatic pressure range using an anvil cell technique. Figure ~\ref{fig5}a) presents the temperature dependence of the tunnel-diode oscillator (TDO) frequency at various pressures with background removed. The background is determined by a linear extrapolation of the data above the superconducting transition to the lowest temperature of a particular run, as depicted in the upper inset to Fig.~\ref{fig5}a). The data is plotted as -1$\times$frequency, so that the superconducting transition appears as a drop in order to facilitate easy comparison with resistivity data. We determine $T_{c}$ by the intersection of two lines as shown in the main panel of Fig.~\ref{fig5}(a). Resistance measurements on Ca$_{0.73}$La$_{0.27}$Fe$_{2}$As$_{2}$, shown in Fig.~\ref{fig5}(b), show complete resistive transitions with $T_{c}$ values that closely follow those observed in the TDO measurement. 

Using the data from both TDO and resistance measurements obtained in the anvil cell, the pressure-temperature phase diagram for overdoped sample S4 has been constructed as shown in Fig.~\ref{fig5}(c). As shown, the SC transition temperature rises from 31~K at ambient pressure to a maximum value of $\sim$ 44~K at 20 kbar. This gives an average slope of $\sim$ +0.65~K/kbar, a value that is considerably higher than other electron-doped 122 pnictides. In overdoped BaFe$_{1.8}$Co$_{0.2}$Fe$_{2}$As$_{2}$ for example, $T_{c}$ was observed to increase with an initial slope of 0.065~K/kbar, \cite{Ahilan} and then level off at higher pressures where $T_{c}$  increases only 1~K when applying 25 kbar. In the underdoped regime, on the other hand, the pressure coefficient does become sizable, reaching 0.4 K/kbar for a sample with composition BaFe$_{1.92}$Co$_{0.08}$Fe$_{2}$As$_{2}$ \cite{Ahilan}. The clear maximum in $T_{c}$ vs. pressure raises the question as to why an overdoped sample of Ca$_{1-x}$La$_{x}$Fe$_{2}$As$_{2}$ displays the pressure sensitivity characteristic of an underdoped sample of BaFe$_{2-x}$Co$_{x}$As$_{2}$.  From a structural perspective, the replacement of Ca (atomic radius 126 pm) with La (130 pm) expands the $a$-axis lattice parameter, from 3.895 \AA ~ for pure CaFe$_{2}$As$_{2}$~ to 3.92 \AA ~ for $x$=0.27 as measured at 250 K. The $c$-axis lattice parameter remains essentially unchanged within the margins of error in the measurement ~\cite{Saha_CaR}. Applying hydrostatic pressure to CaFe$_{2}$As$_{2}$~ has a somewhat different effect, causing a significant shortening of the $c$-axis, from 11.75 \AA ~ to 11.50 \AA ~ by 10 kbar in the high temperature tetragonal phase (T') at room temperature. Over the same pressure range, the $a$-axis lattice constant increases slightly, by 0.1\% ~\cite{Canfield}.
However, it is quite interesting to note that the maximum value of $T_{c}$~ = 44 K seen in our pressure experiments on both under and overdoped Ca$_{1-x}$La$_{x}$Fe$_{2}$As$_{2}$~ samples is very close to that seen at ambient pressure in optimally doped Ca$_{0.8}$La$_{0.2}$Fe$_{2}$As$_{2}$ (maximum $T_{c}$=43 K)~\cite{Saha_CaR, Gao}.


Alternatively, one might consider the role of density fluctuations in
boosting $T_{c}$. For undoped CaFe$_{2}$As$_{2}$~ at room temperature, a
volume collapse into the cT state is known to occur when the interlayer
As-As separation approaches 3.0 \AA ~ ~\cite{Kreyssig,Saha_CaR}. At sufficiently high
pressures it is possible that a cT transition may be induced, as was found in Ca$_{0.67}$Sr$_{0.33}$Fe$_{2}$As$_{2}$~\cite{Jason}.

It is interesting to compare these findings with recent experiments on phosphorous co-doping in the rare earth substituted CaFe$_2$As$_2$ system \cite{Kudo}. Isovalent substitution of P for As ions in the Ca$_{1-x}$La$_x$Fe$_2$(As$_{1-y}$P$_y$)$_2$ appears to create an island of high volume fraction superconductivity, for samples close to the doping levels $x$=0.12 and $y$=0.06. In some pnictides, such as the BaFe2(As$_{1-x}$P$_x$)$_2$ system, the equivalence of P substitution and pressure has been established \cite{Klintberg}. Given this result it seems interesting to probe the superconducting volume fraction in our system of samples with La doping $x$=0.12 under hydrostatic pressure, to see whether a similar island of optimized superconductivity occurs.

\section{General Discussion}


Considering that isovalent P substitution in CaFe$_{2}$As$_{2-x}$P$_{x}$ also induces the structural collapse without a high-$T_{c}$ phase present \cite{Kasahara}, it appears that charge doping is an essential ingredient for the high-$T_{c}$ superconducting phase to stabilize \cite{Saha_CaR}. In Ca$_{1-x}$R$_{x}$Fe$_{2}$As$_{2}$, one proposed scenario is that the low-volume-fraction high-$T_{c}$ phase might be inhomogeneous, originating from local effects tied to the low percentage of rare earth substitution. However, recent STM measurement on Ca$_{1-x}$Pr$ _{x}$Fe$_{2}$As$_{2}$ demonstrated that Pr dopants do not cluster, and in fact show a slight tendency to repel each other at very short length scales~\cite{Hoffman}. These findings suggest that Pr inclusion or its inhomogeneous distribution is unlikely to be the source of the high-$T_{c}$~ phase \cite{Hoffman}, similar to what has been concluded in the case of RFeAsO$_{1-x}$F$_x$ where neutron studies that distinguish between a magnetic Kramer's doublet ground state for R=Nd and a singlet state for R=Pr rule out any direct role of R substitution for the high-$T_c$ phase \cite{Goremychkin}.

Chu \etal\ have suggested the possibility of interfacial superconductivity as the mechanism driving the high-$T_{c}$ phase \cite{Lv2}. In this scenario, superconductivity is occurring at the interface of small cluster-like regions elongated along the $ab$-plane inside the bulk metallic region. 
However, the pressure driven superconductivity observed in the current study does not support the scenario where intergrowth or cluster regions are responsible for the trace high-$T_{c}$ superconductivity. More important, the recent observation of much greater volume fractions of superconductivity in phosphorus-substituted samples~\cite{Kudo} is also inconsistent with such a scenario. 

Rather, our study strongly indicates that long range magnetic order must be fully suppressed in order to induce the high-$T_{c}$~superconducting phase. If cluster or interface regions play any role in stabilizing the high-$T_c$ phase, they must also abruptly appear beyond the AFM-SC quantum phase transition. Given the vastly different rates of suppression of $T_N$ with pressure in non-hydrostatic (Fig.~\ref{fig2}) and hydrostatic (Fig.~\ref{fig32}) pressure experiments, it is highly unlikely that such regions form equivalently under dramatically different strain conditions. It is clear that lattice strain, and the resultant local perturbations on electronic and magnetic structure, must play a key role in stabilizing the superconducting phase, as is well documented in the case of \Sr\ \cite{Saha_Sr} where lattice distortions are directly tied to the appearance of superconductivity. 

Finally, the most striking observations in the Ca$_{1-x}$R$_{x}$Fe$_{2}$As$_{2}$ series involve the scale of the superconducting temperature, reaching near 50~K and approaching that of the highest-$T_c$ oxypnictide 1111 compounds. It is striking that, both in the 1111 materials and in the 122 system considered here, the temperature-doping phase diagrams include a strong segregation between AFM and SC phases with little to no coexistence. (Although coexistence has indeed been observed in the SmFeAsO$_{1-x}$F$_x$ system \cite{Drew}, it is widely thought to originate from a percolative type segragation of AF and paramagnetic-metal domains \cite{Fujiwara}.) It is thus tempting to draw parallels between the origin of the highest-$T_c$ pairing that may be common in the two systems, given these similarities. First, the first-order nature of the separation between AFM and SC phases may have roots in the interplay between magnetic and structural (nematic) instabilities that impinge on the superconducting phase. Theoretical work by Fernandes \etal\ has recently shown that the presence of superconductivity has a profound effect on the first-order character of the quantum phase transition between AFM and paramagnetic/superconducting phases \cite{Fernandes-PRL}, which in the limit of large Fermi surface nesting mismatch drives away the coexistence of the two phases. 
Why $T_c$ is maximized in such systems remains to be understood, but the complete separation of AFM and SC phases is likely to play a role, perhaps due to a lack of anisotropy in the superconducting gap that can derive from the presence of Fermi surface reconstruction in the spin-density wave state \cite{Maiti,Reid}.
Most recently, the emergence of two superconducting domes in the phase diagram of LaFeAsO$_{1-x}$H$_x$ \cite{FujiwaraPRL} is a tantalizing hint at the possibility of two distinct superconducting phases that may find a basis in these closely related but distinct iron-based materials.

In summary, we have employed both quasi-hydrostatic and hydrostatic pressures to illuminate the pressure-temperature phase diagrams of underdoped Ca$_{1-x}$La$_{x}$Fe$_{2}$As$_{2}$ as well as the quasi-hydrostatic pressure dependence of an overdoped sample of Ca$_{1-x}$La$_{x}$Fe$_{2}$As$_{2}$ with $x$=0.27. Interestingly, high-temperature superconductivity in the Ca$_{1-x}$R$_{x}$Fe$_{2}$As$_{2}$ series appears after the complete suppression of the AFM phase, with little or no coexistence between the two phases. This is strikingly similar to the segregation of SC and AFM phases found in 1111 materials doped with fluorine, such as in LaFeAsO$_{1-x}$F$_{x}$  \cite{Zhao}, CeFeAsO$_{1-x}$F$_{x}$  \cite{Luetkens}, and SmFeAsO$_{1-x}$F$_{x}$ \cite{Maeter} and should be contrasted with the well-known coexistence shown to occur in BaFe$_{2-x}$Co$_{x}$As$_{2}$ ~\cite{Ni,Chu,Klintberg}. Complete suppression of magnetic order appears to be a necessary condition for the high-$T_{c}$~ superconductivity as in the case of the 1111 family. In the overdoped sample, a rapid rise in $T_{c}$~ where one might naively expect suppression suggests that the application of pressure in this system does not simply follow the expectation from doping, and offers an additional route to tune $T_{c}$~ in this family of materials. 
Above all, the unusual dichotomy between lower-$T_{c}$~ systems that happily coexist with antiferromagnetism and the tendency for the highest-$T_{c}$~ systems to show no coexistence is an important clue to the pairing mechanism in iron-based superconductors.


The authors acknowledge R.L. Greene, P. Canfield, G. Lonzarich and P. Alireza for useful discussions, and thank J. Leao for assistance with neutron diffraction measurements under pressure. Research at the University of Maryland was supported by the AFOSR-MURI (FA9550-09-1-0603) and the NSF (DMR-0952716). M.S. acknowledges support from the Royal Society, and S.K.G. from Trinity College Cambridge. 

\end{document}